\documentclass[11pt, oneside]{article}   	
\usepackage{geometry}                		
\usepackage{graphicx,color}				
\usepackage{amssymb}
\usepackage{authblk}
\usepackage{blindtext}

\def\comm#1{\textcolor{black}{#1}}

\def\ka{K_{1a}}
\def\kb{K_{1b}}
\def\kL{K_{1}(1270)}
\def\kH{K_{1}(1400)}
\def\tK{\theta_{K_1}}
\def\tzero{\theta_0}
\def\mixing{\theta_{K_1}}

\def\eq#1{Eq.~(\ref{#1})}
\def\eqs[#1]#2{Eqs.~(\ref{#1})-(\ref{#2})}

\def\gev{{\rm\ GeV}}
\def\mev{{\rm\ MeV}}

\def\lsim{\raise0.3ex\hbox{$\;<$\kern-0.75em\raise-1.1ex\hbox{$\sim\;$}}}
\def\gsim{\raise0.3ex\hbox{$\;>$\kern-0.75em\raise-1.1ex\hbox{$\sim\;$}}}

\def\fivebodydecay{$\tau^-\to K_1^-\nu_\tau \to (K^-\omega) \nu_\tau\to (K^- \pi^+\pi^-\pi^0)\nu_\tau \ $}

\newcommand{\be}{\begin{equation}}
\newcommand{\ee}{\end{equation}}
\newcommand{\bea}{\begin{eqnarray}}
\newcommand{\eea}{\end{eqnarray}}

\title{The hadronic $\tau$ decay \fivebodydecay and the axial vector mixing angle}
\author[1]{K.~Hayasaka} 
\author[2]{Z.~Huang}
\author[2]{E.~Kou}
\affil[1]{Niigata University, 8050 Ikarashi 2-no-cho, Nishi-ku, Niigata, 950-2181, Japan }
\affil[2]{Universit\'{e} Paris-Saclay, CNRS/IN2P3, IJCLab, 91405 Orsay, France}
\date{}							

\begin{document}
\maketitle
\vspace*{1.2cm}
\begin{abstract}
We propose to measure the \fivebodydecay decay in order to determine the $K_1$ axial vector mixing angle $\mixing$. 
We  derive, for the first time, the differential decay rate formula for this decay mode. Using the obtained result, 
we perform a sensitivity study  for the Belle (II) experiment. We will show that the $K^-\pi^+\pi^-\pi^0$ spectrum of the  \fivebodydecay decay can {discriminate}  the two solutions $\mixing=\sim 30^{\circ}$ or $\sim 60^{\circ}$ observed in \comm{the} other measurements. 
\end{abstract}

\newpage
\section{Introduction}
The hadronic $\tau$ decay is a very useful tool to investigate the nature of the light hadrons. 
The initial state being lepton allows to study the strong decays of the final state hadrons in a clean manner. 
The hadrons \comm{being} produced from the $W$ boson provides a valuable information on the vector and the axial vector couplings of the hadrons.  
In this article, we investigate the $\Delta S=1$ hadronic $\tau$ decay. This type of decays is Cabibbo suppressed but offers unique way to explore the nature of the Kaonic resonances~\cite{Kou:2018nap}.  
We investigate the \fivebodydecay decay to obtain the information of the $K_1$ axial vector mesons, $\kL$ and $\kH$. A better understanding of the $K_1$ mesons is not only an interest of its own but also is highly demanded  by $B$ physics recently. In $B$ physics, to disentangle the new physics effect from the hadronic uncertainties is the {essential task}  for a discovery. The recent studies of $B\to K_1\gamma$ decays~\cite{Kou:2010kn, Bellee:2019qbt, Kou:2013gna, Wang:2019wee, Adolph:2018hde}, $B\to K_1 \ell^+\ell^-$~\cite{Huang:2018rys} or $B\to K_1\pi$ decays~\cite{Cheng:2007mx, Cheng:2008gxa, Dalseno:2019kps}, which are known to be sensitive to the new physics  coming from the right-handed current or the CP violation, respectively, show that a more accurate information of the $K_1$ mesons would enhance the sensitivity to the new physics. 

In this article, we propose \comm{to measure} the \fivebodydecay decay to determine the  $\mixing$ angle. The $\mixing$ enters both in the production and the decay of $K_1$ meson in this process. The determination of the axial-vector mixing angle caused a controversy.  Mainly two ways to determine $\mixing$  have been attempted, i) mass fit assuming the $SU(3)$, ii) strong decay of $K_1$. Both show basically two possible solutions far apart, around $\sim 30^{\circ}$ and $\sim 60^{\circ}$ (see e.g.~\cite{Suzuki:1993yc, Burakovsky:1997dd, Li:2005eq, Tayduganov:2011ui, Cheng:2011pb, Cheng:2013cwa}). 
In this article, we  show  the result of the 5 body differential decay rate, \fivebodydecay, for the first time. Then, we use this result to perform a sensitivity study \comm{for  $\mixing$ determination} at the Belle II experiment. This process was studied in ALEPH~\cite{Buskulic:1996qs,Barate:1999hj}  and  CLEO~\cite{Arms:2005qg} experiments and a few hundreds of events are observed. The Belle II experiment can \comm{acquire} 2-3 orders of magnitudes more data in the future. 

The remaining of the article is organised as follows. In section 2, we derive the 5 body differential decay rate. In section 3, we introduce the mixing angle and rewrite our results in terms of $\mixing$. We show our numerical result and the Monte Carlo study assuming the Belle (II) setup in section 4 and we conclude in section 5.

\section{Differential Decay Rate of \fivebodydecay }
We first present the computation for the decay rate of the five body decay {(4 momentum  associated to each particle is given in the parenthesis)}
\be
\tau^- (Q)\to K_1^- (k_1)\nu_{\tau}(p_0) \to K^-(p_1) \omega(k_2)\nu_{\tau}(p_0) \to K^-(p_1) \pi^+(p_2) \pi^-(p_3) \pi^0(p_4)\nu_{\tau}(p_0)
\ee
where $K_1$ is a $J^{P}=1^+$ meson, i.e. $K_1(1270)$ or $K_1(1400)$. 
The \comm{five body} differential decay rate can be given as 
\be
{d\Gamma} = \frac{(2\pi)^4}{2m_\tau} |{\mathcal{M}}|^2d\Phi_5 
\ee
where 
\be
d\Phi_5
= \frac{1}{(2\pi)^{14}} \frac{1}{(2^7m_\tau \sqrt{k_2^2})}|\vec{\tilde{p}}_1||\vec{p}_0|
 {d\sqrt{k_2^2}d\sqrt{k_1^2}}
 dm_{23}^2dm_{34}^2  d (\cos\bar{\theta}) d\bar{\phi} d\bar{\psi} {d\tilde{\Omega}}  d\Omega 
\ee
{The variables with $\tilde{}$ and $\bar{}$ are the momentum and polar angles in the rest frame of $K_1$ and $\omega$, respectively. }
Since the \comm{angular dependence is not easy to measure} in $\tau$ decays, we integrate them all \comm{in this work}.
Thus, {the remaining} integration variables are  the invariant masses of $K_1, \omega$ and two Dalitz variables of $\omega$ decays, which are given as
\be
k_1^2=(p_1+p_2+p_3+p_4)^2, \quad k_2^2=(p_2+p_3+p_4)^2, \quad {m_{23}^2}=(p_2+p_3)^2, \quad {m_{34}^2}=(p_3+p_4)^2
\ee
{The 3-momentum of $\nu_\tau$, $|\vec{p}_0|$, and of the final state $K$, $|\vec{\tilde{p}}_1|$, are written by the integration variables, $\sqrt{k_2^2}$ and $\sqrt{k_1^2}$, as}
\bea
|\vec{p}_0| &=& {\frac{{m_{\tau}^2-{k_1^2}}}{2m_{\tau}}}\label{Eq:32}\\
|\vec{\tilde{p}}_1| &=& \frac{\sqrt{\left(k_1^2-(m_1+\sqrt{ k_2^2})^2\right)\left(k_1^2-(m_1-\sqrt{ k_2^2})^2\right)}}{2\sqrt{k_1^2}} \label{Eq:33} 
\eea

The decay amplitude $\mathcal{M}$ is obtained as a product of the {successive} decay amplitudes, i.e.: 
\be
{\mathcal{M}}={\mathcal{M}}_3(\omega\to \pi^+\pi^-\pi^0)\times 
{\mathcal{M}}_2(K_1^-\to K^-\omega)\times
{\mathcal{M}}_1(\tau^- \to K_1^-\nu_\tau )
\ee

The amplitude of the $\tau\to K_1\nu_\tau$ can be written as
\be
{\mathcal{M}}_1(\tau \to K_1\nu_\tau )=\frac{G_F}{{m_{\tau}}}V_{us}^* j_\mu \langle K_1 | \overline{s}\gamma^\mu (1-\gamma_5)u |0\rangle
\ee
where the leptonic current is given as
\be
j_\mu=\overline{\nu}_\tau \gamma_\mu (1-\gamma_5) \tau 
\ee
The $K_1$ meson can be produced only from the axial vector current and the matrix element of $K_1$ production  is given by {a decay constant $f_{K_1}$}  
\be
\langle K_1 | \overline{s}\gamma^\mu (1-\gamma_5)u |0\rangle ={-i f_{K_1} m_{K_1}}\epsilon^{*\mu}(k_1)
\ee
where $K_1$ is only symbolic here and it can mean $K_1(1270)$ or $K_1(1400)$. The detailed definitions of the decay constants for these two states are given in the next section. 

The amplitude of the $K_1\to K\omega$ decay can be written by the two form factors 
\be
{\mathcal{M}}_2(K_1\to K\omega)=\epsilon_{K_1}^\mu T_{\mu\nu}\epsilon_{\omega}^{*\nu}
\ee
where 						
\be
T_{\mu\nu}=f^{K_1}g_{\mu\nu}+h^{K_1} k_{2\mu} p_{1\nu}
\ee
Note that these form factors can be related to the S-wave and P-wave amplitudes (see Appendix D of~\cite{Tayduganov:2011xna} for derivation)
\bea
f^{K_1}&=& -A^{K_1}_S-\frac{1}{\sqrt{2}}A^{K_1}_D \\
h^{K_1}&=& \frac{\tilde{E}_\omega}{\sqrt{k_1^2}|\vec{\tilde{p}}_1|^2}\left[\left(1-\frac{\sqrt{k_2^2}}{\tilde{E}_\omega }\right)A^{K_1}_S+{\left(1+\frac{2\sqrt{k_2^2}}{\tilde{E}_\omega }\right)}\frac{1}{\sqrt{2}}A^{K_1}_D\right]
\eea
where $\tilde{E}_\omega=\sqrt{|\vec{\tilde{p}}_1|^2+k_2^2}$. 
As the decay rates of S-wave and D-wave are not separately known, we must rely on the theoretical model {as we will see late-on}.  

The amplitude of the $\omega\to \pi^+\pi^-\pi^0$ can be written by one form factor
\be
{\mathcal{M}}_3(\omega\to \pi^+\pi^-\pi^0)=ig \epsilon _{\mu\nu\alpha\beta}\epsilon^\mu p_2^\nu p_3^\alpha p_4^\beta {\mathcal{F}}
\ee
where assuming that the $\omega \to 3\pi $ go through three possible resonances, $\rho^+, \rho^-$ and $\rho^0$, we can simply write the form factor to be
\be
{\mathcal{F}}= \frac{1}{{m_{24}^2}-m_{\rho^+}^2+im_{\rho^+} \Gamma_{\rho^+}}+\frac{1}{{m_{34}^2}-m_{\rho^-}^2+im_{\rho^-} \Gamma_{\rho^-}}+\frac{1}{{m_{23}^2}-m_{\rho^0}^2+im_{\rho^0} \Gamma_{\rho^0}}
\ee
where {we assign $p_{2,3,4}$ as the 4-momentum of $\pi^{+,-,0}$. }

{Finally, } the squared amplitudes after integration of all the angles is obtained as\footnote{As we integrate all the angles, all the spins can be summed after squaring the amplitude.} 
\be
\frac{d\Gamma(\tau\to K_1\nu \to K\omega \nu\to K\pi\pi\pi\nu)}{d\sqrt{k_1^2}d\sqrt{k_2^2}dm_{23}^2dm_{34}^2} \\
 = {\frac{(2\pi)^4}{2m_\tau}}|\mathcal{M}|^2\frac{1}{(2\pi)^{10}} \frac{1}{(2^4m_\tau \sqrt{k_2^2})}|\vec{\tilde{p}}_1||\vec{p}_0| \nonumber 
 \ee
\be
|\mathcal{M}|^2 =
 {\left(\frac{G_F V_{us}m_{K_1}g }{m_\tau}\right)^2 |\mathcal{F}|^2}\times \frac{512\pi^4}{27k_1^2}m_\tau  |\vec{p}_0||\vec{\tilde{p}}_2|^2|\vec{\tilde{p}}_3|^2\sin^2\delta ( 4 m_\tau^2+12 k_1^2 + \frac{2 k_1^4}{m_\tau^2})\mathcal{C}
\ee
where 
\[{\delta=\cos^{-1}\Big[\frac{E_{\pi^+} E_{\pi^-}+\frac{2m_{\pi}^2-m^2_{23}}{2}}{\sqrt{E_{\pi^+}^2-m_\pi^2}\sqrt{E_{\pi^-}^2-m_\pi^2}}\Big].}\]
The factor $\mathcal{C}$ is 
\[\mathcal{C}=(2|F_0|^2+|F_1|^2) \] 
with 
\bea
F_0 &=& f_{K_1}f^{K_1}\mathcal{BW}_{K_1} \\
F_1&=& \Big[f_{K_1}(f^{K_1}\tilde{E}_\omega+h^{K_1}\sqrt{k_1^2}|\vec{\tilde{p}}_1|^2)\mathcal{BW}_{K_1}\Big]/\sqrt{k_2^2}\nonumber \\
\eea
and 
\[{\mathcal{BW}_{K_1} =\frac{1}{k_1^2-m_{K_1}^2+im_{K_1} \Gamma_{K_1}}}\]
As we are interested in the contributions from $K_1 (1270)$ and $K_1 (1400)$ as well as their interference, we sum them at the amplitude level and take a square. Further replacing the form factors to the partial wave amplitudes, we obtain the $\mathcal{C}$ function  as 
\bea
\mathcal{C}&=& 3 \Big\{
\Big| f_{K_1(1270)}A_S^{K_1(1270)}{\mathcal{BW}}_{K_1(1270)}+ f_{K_1(1400)}A_S^{K_1(1400)}{\mathcal{BW}}_{K_1(1400)}\Big|^2   \label{eq:2-21}\\
&&\quad \quad \quad \quad +\Big| f_{K_1(1270)}A_D^{K_1(1270)}{\mathcal{BW}}_{K_1(1270)}+ f_{K_1(1400)}A_D^{K_1(1400)}{\mathcal{BW}}_{K_1(1400)}\Big|^2\Big\}  \nonumber
\eea

In the next section, we obtain the decay constants $f_{K_1 (1270,1400)}$ as well as the partial wave amplitudes $A_{S,D}^{K_1(1270,1300)}$ in terms of the axial vector mixing angle $\mixing$.

\section{The axial vector mixing angle $\mixing$}
The axial vector strange mesons have a  peculiar nature, the observed physical states, $\kL$ and $\kH$ are the mixture of two $J^{P}=1^{+}$ states, $^3P_1$ and $^1P_1$.
This is different from the case of the non-strange axial vector mesons,  $a_1(1260)$ and $b_1(1235)$, which do not mix as the $^3P_1$ and $^1P_1$ states are also the eigenstates of  different intrinsic charge, {i.e.  ($J^{PC}=1^{++}, 1^{+-}$).} 
Let us denote the unphysical $^3P_1$ and  $^1P_1$ states as $\ka$ and $\kb$, respectively. 
Then, the the physical states (mass eigenstates) can be written  as
\begin{equation}
\left(\begin{array}{c}\kL\\ \kH \end{array}\right)
=\left(\begin{array}{cc}\sin \tK & \cos \tK \\ \cos \tK & -\sin \tK \end{array}\right)
\left(\begin{array}{c}\ka\\ \kb \end{array}\right)
\label{eq:su3-1}
\end{equation}
where  $\tK$ is called as the axial vector mixing angle. 

Investigating the nature of the strange axial vector mesons produced from $\tau$ decay, i.e. the weak interaction, has a great advantage. 
The spin singlet configuration of the $s$ and $u$ quarks are suppressed with respect to the spin triplet one as the former is chirally forbidden and furthermore, the $SU(3)$ and charge symmetry forbids the production of $\kb$~\cite{Lipkin:1992tw}. This leads to, at the first order, that only the $\ka$ is produced from the weak interaction. Therefore, by defining the decay constant of $\ka, \kb$ state, 
\be
\langle K_{1a, 1b}  | \overline{s}\gamma^\mu (1-\gamma_5)u |0\rangle ={-i} f_{K_{1a, 1b} }{m_{K_{1a, 1b}}} \epsilon^{*\mu}(k_1), 
\ee
we have $f_{\ka}\gg f_{\kb}\simeq 0$. 
Then, by using \eq{eq:su3-1}, the decay constant of the physical states can be given as 
\be
f_{\kL}= f_{\ka} \sin \mixing, \quad  f_{\kH}= f_{\ka} \cos \mixing   
\ee
Since the $s$ quark mass is not completely negligible with respect to the $K_1$ masses,  $f_{\kb}$ {may not  vanish}. This effect can be  taken into account by  shifting the mixing angle by  $\delta_s= \tan^{-1}\Big(\frac{f_{\kb}}{f_{\ka}}\Big)$,
\be
f_{\kL}= f_{\ka} \sin \mixing^{\prime}, \quad  f_{\kH}= f_{\ka} \cos \mixing^{\prime}  
\ee
where $\mixing^{\prime}\equiv \mixing +\delta_s$. {We investigate maximum of }  10 \% of $s$ quark mass effect, {i.e.} $|\delta_s|<0.1$ ($6^{\circ}$), {in the following}. 

Next, we consider the strong decay, $K_1\to K\omega$. We use the result of the quark model computation in~\cite{Tayduganov:2011ui}, where a similar process, $K_1\to K\rho$ decay, is investigated. 
Using the $SU(3)$ symmetry, the $S$-wave and $D$-wave amplitudes for the $\ka$ and $\kb$ states can be written by  the universal amplitudes, $S^{ABC}$ and  $D^{ABC}$ as (see~\cite{Tayduganov:2011ui} for derivation)
\be%
A_S^{\ka}=\sqrt{\frac{2}{3}}S^{ABC},   \quad A_D^{\ka}=-\frac{1}{\sqrt{3}}D^{ABC}, \quad
A_S^{\kb}=\frac{1}{\sqrt{3}}S^{ABC},   \quad A_D^{\kb}=\sqrt{\frac{2}{3}}D^{ABC}. 
\ee
Then, the amplitudes for the physical states yield 
\bea
A_S^{K(1270)} &=& A_S^{\ka} \sin\theta_{K_1} + A_S^{\kb} \cos \theta_{K_1} = S^{ABC}\sin(\mixing+\tzero)\\
A_D^{K(1270)} &=& A_D^{\ka} \sin\theta_{K_1} + A_D^{\kb} \cos \theta_{K_1} =D^{ABC}\cos(\mixing+\tzero)\\
A_S^{K(1400)} &=& A_S^{\ka} \cos\theta_{K_1} - A_S^{\kb} \sin \theta_{K_1}=-S^{ABC} \cos(\mixing+\tzero)\\
A_D^{K(1400)} &=& A_D^{\ka} \cos\theta_{K_1} - A_D^{\kb} \sin \theta_{K_1}=-D^{ABC} \sin(\mixing+\tzero)
\eea
where $\tzero=\tan^{-1}\frac{1}{\sqrt{2}}\simeq 35.26^{\circ}$. It is important to mention that we do not expect a $SU(3)$ breaking effect beyond this result.
It is because  {the $\ka$ and $\kb$ mixing occurs via the very hadronic decays,  $K_1\to K\omega$ as well as $K\rho, K^*\pi$} (i.e. hadronic contributions in the loop). {The $SU(3)$ breaking effect, which comes from the mass difference among these intermediate states, is taken into account via the non-zero $\mixing$ angle. }

Finally, we can simplify Eq.~(\ref{eq:2-21}) by using the mixing angle as 
\bea
\mathcal{C}&=& 3 |f_{\ka}|^2\Big\{|S^{ABC}|^2\Big|\sin\theta_{K_1}^{\prime}\sin(\theta_{K_1}+\theta_0){\mathcal{BW}}_{K_1(1270)}+\cos\theta_{K_1}^{\prime}\cos(\theta_{K_1}+\theta_0){\mathcal{BW}}_{K_1(1400)} \Big|^2 \nonumber \\
&&\quad  \quad \quad +|D^{ABC}|^2\Big|\sin\theta_{K_1}^{\prime}\cos(\theta_{K_1}+\theta_0){\mathcal{BW}}_{K_1(1270)}+\cos\theta_{K_1}^{\prime}\sin(\theta_{K_1}+\theta_0){\mathcal{BW}}_{K_1(1400)}
\Big|^2\Big\} \nonumber\\ 
\label{eq:3-32}
\eea
which {is our final result and will be used in the next section}. 
It should be emphasised that the obtained expression is different from the one proposed in~\cite{Suzuki:1993yc}, where only the $\mixing$ dependence on the $\tau\to K_1\nu$ decay is taken into account but not on the $K_1$ decay. 

\section{The numerical results and Belle (II) sensitivity study}
In this section, we present the sensitivity of the Belle II experiment to the $\mixing$ angle. First, we list up all the parameters we use in our numerical analysis. Note that for now, we list only the central values while we will discuss the uncertainties associated to them later-on: 
\bea
& m_\tau = 1.777\gev, \quad m_{\kL}=1.270\gev, \quad m_{\kH}=1.400\gev & \nonumber \\
& m_K=0.494\gev, \quad m_\pi=0.135\gev, \quad m_\omega=0.782\gev, \quad m_\rho=0.775\gev  & \label{eq:4-33}\\
&\Gamma_{\kL}=0.09\gev, \quad \Gamma_{\kH}=0.174\gev, \quad \Gamma_\omega= 0.00849\gev, \quad \Gamma_\rho=0.148\gev& \nonumber 
\eea
Since our goal is not to estimate the total decay rate, the \comm{overall} factors {not listed her, such as $G_F, V_{us}, g, f_{K_1}\cdots$ are not necessary in this study}. 
For the universal partial wave amplitude, which we introduced in the previous section,  the result from the $^3P_0$ model~ yields~\cite{Tayduganov:2011ui}
\be
S^{ABC}\propto (3-\alpha |\vec{\tilde{p}}_1|^2)e^{-\beta|\vec{\tilde{p}_1}|^2}e^{-f_2 (|\vec{\tilde{p}}_1|^2-|\vec{\tilde{p^0}_1}|^2)}, \quad 
D^{ABC}\propto \alpha |\vec{\tilde{p}}_1|^2e^{-\beta|\vec{\tilde{p}}_1|^2}e^{-f_2 (|\vec{\tilde{p}}_1|^2-|\vec{\tilde{p}^0}_1|^2)}
\ee
where {$\alpha=4.2 \gev^{-2}, \beta=0.52 \gev^{-2}$} and $f_2=3.0$. The last exponential  is the so-called damping factor, which introduces the cut off for the large momentum region. The momentum $\vec{\tilde{p}^0}_1$ is  $\vec{\tilde{p}}_1$ at the pole masses. For $\kL$, there is no available phase space at the pole mass, thus, the damping factor can be neglected. 

In order to have an idea of the $D$-wave contribution, let us quote the mean values of $\vec{\tilde{p}}_1$ for $\kL$ and $\kH$
\be
\langle \vec{\tilde{p}}_1 \rangle_{\kL}= (0.19 \pm 0.09)\gev, \quad 
\langle \vec{\tilde{p}}_1 \rangle_{\kH}= (0.28 \pm 0.09)\gev
\ee
where the error comes from the spread of the $\langle \vec{\tilde{p}}_1 \rangle$. This number implies that the $D$-wave amplitude is roughly 5(10)\% for $\kL(\kH)$ of the $S$-wave one. As these two contributions do not interfere, we expect the $D$-wave contributions is very small. 
In order to simplify the analysis, we use a constant $S^{ABC}$ and $D^{ABC}$ in the present study. 
So as to take into account the momentum depending term in the $S$-wave, which is not negligible,  we choose different constants for $\kL$ and $\kH$. We find the following choices reproduce well the full expression: 
\be
{S^{ABC}_{\kL} = 3.0 \gev, \quad D^{ABC}_{\kL} = 0.2 \gev, \quad 
S^{ABC}_{\kH} = 2.3 \gev, \quad D^{ABC}_{\kH} = 0.3 \gev} \label{eq:36}
\ee
which we will use in our analysis. Later in this section, we discuss the impact of the variations of these parameters. 

\begin{figure}[htbp]
\begin{center}
\includegraphics[width=10cm]{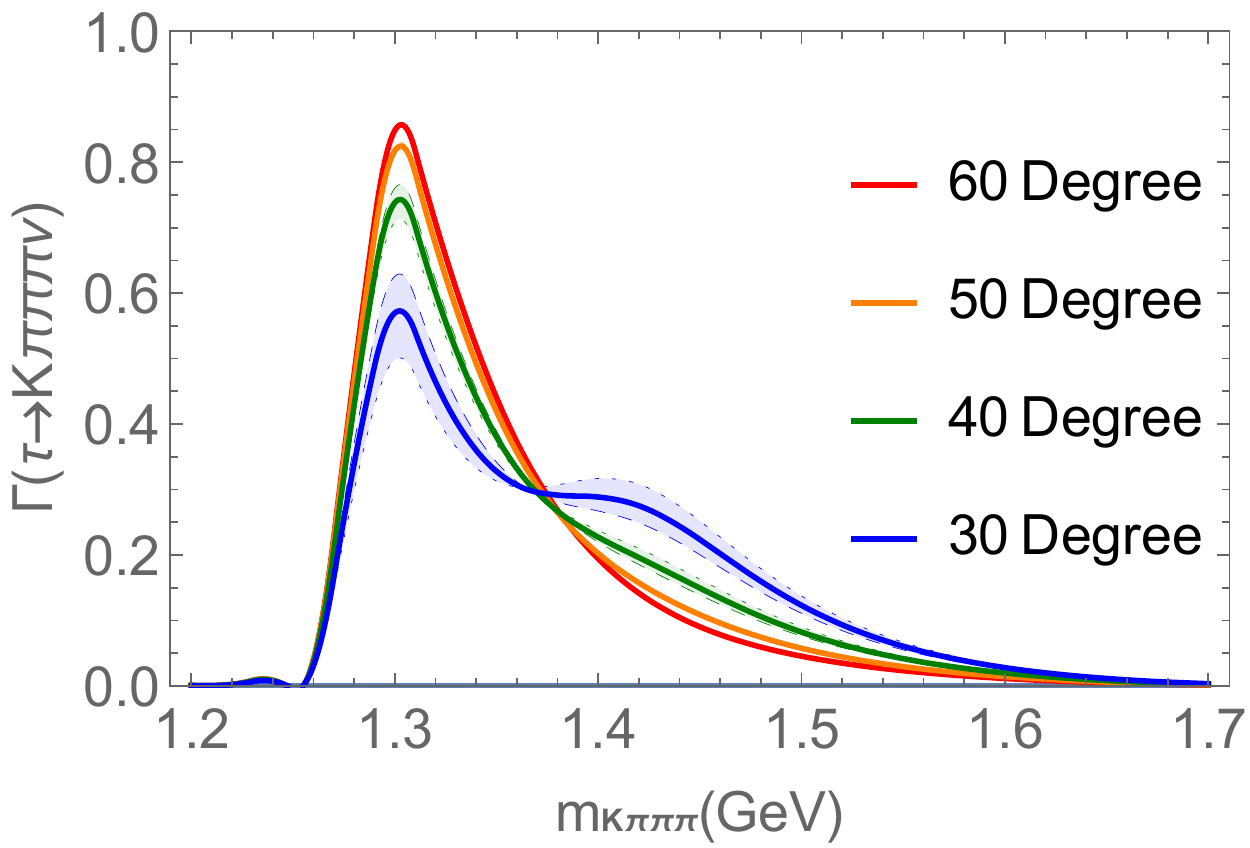}
\caption{The $K\pi\pi\pi$ distribution of the \fivebodydecay. The solid line represents the result without the extra $SU(3)$ breaking effect (see text), i.e. $\delta_s=0$ while the coloured area is with this effect  with amount of $|\delta_s|\lsim 6^{\circ}$ (the dashed lines are the results for $\delta_s=6^{\circ}$ and the dotted lines are for $\delta_s=-6^{\circ}$). }
\label{fig:1}
\end{center}
\end{figure}

The $K\pi\pi\pi$ mass distribution (normalised to unity) is shown in Fig.~\ref{fig:1}. The solid line is the result with no  extra $SU(3)$ breaking, mentioned earlier, i.e. $\mixing^{\prime}=\mixing$. 
For the small values of $\mixing$, let's say around 30$^{\circ}$, the $K\pi\pi\pi$ spectrum changes significantly for a  variation of the mixing angle while when the mixing angle reaches around $\sim 50^{\circ}$, $\kL$ becomes totally dominant and it becomes difficult to distinguish the results with different $\mixing$. This pattern can be readily inferred    {from  the dominant $S$-wave contributions in} Eq.~(\ref{eq:3-32}). 
The coefficient for the $\kL$ contribution, $\sin\mixing \sin(\mixing +\tzero)$, is an increasing function in the region of $\mixing$ we are considering. On the other hand, the coefficient for $\kH$,  $\cos\mixing \cos (\mixing +\tzero)$, rapidly decreases  and hits zero at $\mixing=90^{\circ}-\tzero=54.74^{\circ}$. 
The coloured bound in Fig.~\ref{fig:1} is results including the extra $SU(3)$ breaking effect with amount of $|\delta_s|\le 6^{\circ}$. {We can see that this effect has an impact} only on the  $\sin\mixing$ and $\cos \mixing$ terms, and as a result, it is almost negligible for $\mixing \gsim 40^{\circ}$. 

In order to clarity the achievable limit by the Belle (II) experiment, we perform a Monte Carlo study. 
The $e^+e^-\to \tau^+\tau^-$ process is simulated by using the KKMC package~\cite{Jadach:1999vf,Jadach:2000ir} with the Belle beam energy, 8\gev \ for electron and 3.5\gev \ for positron. 
We decay the tagging side of $\tau$ by using the TAUOLA package~\cite{Jadach:1990mz,Jezabek:1991qp,Jadach:1993hs}. We do not consider the spin correlation as we will use only the leptonic decay ($e$ or $\mu$) on the tagging side, which reduces significantly the $q\bar{q}$ background.  For the signal side, we use the differential decay rate formulae derived in this article to generate the \fivebodydecay decay distribution. 

The main background comes from $\tau \to 3\pi \pi^0\nu$ decay, where $3\pi$ does not necessarily come from $\omega$ but all possible intermediate states, such as $a_1\pi, \rho\rho, \cdots$. Thus, we select  the four charged tracks with no net charge and evaluate the thrust axis. Here, the `good' charged tracks are defined as $dr<5 mm, |dz|<5cm, p_t>100 \mev$ and `good' gamma is the one with $E_{\gamma}>50$ MeV within the detector fiducial volume. We select the events which have three charged tracks parallel to the thrust axis (signal side) and one anti-parallel (tag side). The signal side should contain two $\gamma$ with $120{\mev}<M_{\gamma\gamma}<150\mev$ and $\pi\pi\gamma\gamma$ in the $\omega$ mass region, $760\mev<M_{\pi\pi\gamma\gamma}<800\mev$. We select only those {$\gamma\gamma$} in the barrel region to avoid the background. For the simplicity, we ignore the multi-candidate case: if more than two sets of $\gamma$ satisfy the above condition, we reject such events (the fraction is around a few percents). The charged track which does not construct $\omega$ is considered to be kaon. For the detection efficiency computation, we use the Belle detector simulation {with the improved kaon identification (ID)}. {That is, we assume kaon ID of Belle II,  90 \% for kaon ID and 4\% for $\pi$ fake rate for kaon ID~\cite{Kou:2018nap}, which is about twice better than Belle.} 

We found that the detection efficiency is {1-2\%}, which results in $\sim$10k event for each $\sim$1 ab$^{-1}$ of data. {Thus, this amount of data is} already available in the Belle experiment. We use 15k event as a benchmark experimental setup in the following analysis. 

\begin{figure}[htbp]
\begin{center}
\includegraphics[width=10cm]{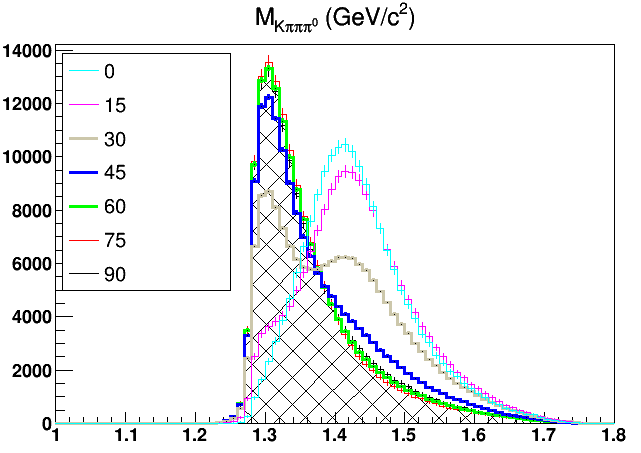}
\caption{The $K\pi\pi\pi$ invariant mass distribution of the \fivebodydecay after taking into account the detector effect of Belle. }
\label{fig:2}
\end{center}
\end{figure}

For the Belle (II) sensitivity study to the axial vector mixing angle $\mixing$, we first generate  events for different values of $\mixing$ from 0 to 90 degree. We use the same input parameters as Fig.~\ref{fig:1}.  The $M_{K\pi\pi\pi}$ spectrum after taking into account the detector effects is given in Fig.~\ref{fig:2}. The similar spectrums are observed in  Fig.~\ref{fig:1} and  Fig.~\ref{fig:2}, which show that our selection criteria is appropriate.

Next, using the generated events, we fit the the $\mixing$ angle. With the 15k of events, we find the statistical  error to be 
\bea
&\sigma_{(\mixing=15^{\circ})}= 0.3^{\circ}, \quad  \sigma_{(\mixing=30^{\circ})}=0.2^{\circ},  &  \\
& \sigma_{(\mixing=45^{\circ})}=0.4^{\circ}, \quad \sigma_{(\mixing=60^{\circ})}=1.1^{\circ}, \quad \sigma_{(\mixing=75^{\circ})}=1.9^{\circ}&\nonumber
\eea
The very small errors estimated with the amount of data which will be soon available, are very encouraging. Thus, we further investigate the various systematic uncertainties. This is particularly important as the $K\pi\pi\pi$ invariant mass distributions for  $\mixing \gsim 40^{\circ}$ seem to be very difficult to distinguish and  the systematic effect could dominate.  To evaluate the experimental systematic error is beyond the scope of this article while we investigate the systematic errors coming from the input parameters in the following.

 As mentioned earlier, since our goal is to determine the mixing angle and  not  the total branching ratio, the overall factors do not induce an uncertainty.  
 The most uncertain input parameters are the mass and width of the $\kL$ resonance in Eq.~(\ref{eq:4-33}) as well as the $S$-wave and $D$-wave amplitudes in Eq.~(\ref{eq:36}). Both induce the similar kinds of uncertainties in the line shape of the $K_1$ resonances.  To identify the line shape of the $K_1$ resonance is a long-standing challenge: the dominant decay channel $\kL \to \rho K$ has no phase space at the pole mass, which distorts the line shape~\cite{EKFLD}.  This issue is investigated intensively in~\cite{Tayduganov:2011ui}  using the kaon beam experiment data~\cite{Daum:1981hb}. Our prescription, to take into account the line shape ambiguity, here is that we free the mass and width of $\kL$ while fitting the $\mixing$. We emphasise that this prescription can accommodate not only the mass and width uncertainties but also the uncertainties induced by the model parameters, the $S$-wave and $D$-wave amplitudes.
Our result for 15k event yields, 
\bea
&\sigma_{(\mixing=15^{\circ})}= 1.3^{\circ}, \quad  \sigma_{(\mixing=30^{\circ})}=1.4^{\circ},  &  \label{eq:38}\\
& \sigma_{(\mixing=45^{\circ})}=1.3^{\circ}, \quad \sigma_{(\mixing=60^{\circ})}=2.6^{\circ}, \quad \sigma_{(\mixing=75^{\circ})}=8.2^{\circ}&\nonumber
\eea
 The fitted mass and width are well within their uncertainties, i.e. $(1.270\pm 0.006)\gev$ and $(0.090\pm 0.013)\gev$, respectively. This clearly shows  the difficulty of determining the $\mixing$ angle at a few degree precision above $\sim 45^{\circ}$. However, it is quite faire to say that the \fivebodydecay decay has certainly an ability to discriminate the two solutions, $\mixing\sim 30^{\circ}$ and $\mixing\sim 60^{\circ}$,  obtained by the other experiments. 
 
Next, we study a possible systematic uncertainty caused by the $SU(3)$ breaking effect.  We discuss this systematic effect separately here, since, as mentioned earlier, the existence of the $SU(3)$ effect is still debatable: more theoretical investigation is needed to clarify whether this effect must be taken into account or not. 
We estimate the $SU(3)$ breaking effect as follows.  We perform a fit of the same 15k event sample by introducing non-zero $\delta_s$ and measure the shift of the $\mixing$ value. In order to estimate the maximum effect,  we vary $\delta_s$ maximally, i.e. $\delta_s=-6^{\circ}(+6^{\circ})$. 
The obtained results are (mass and width are fitted simultaneously)
 \bea
&\Delta{\mixing}_{(\mixing=15^{\circ})}= +3.8^{\circ}(-3.7^{\circ}), \quad  \Delta{\mixing}_{(\mixing=30^{\circ})}=+2.8^{\circ}(-2.8^{\circ}),  &  \\
& \Delta{\mixing}_{(\mixing=45^{\circ})}=+1.4^{\circ}(-1.6^{\circ}), \quad \Delta{\mixing}_{(\mixing=60^{\circ})}=+1.7^{\circ}(+4.7^{\circ}), &\nonumber \\
& \Delta{\mixing}_{(\mixing=75^{\circ})}=-7.7^{\circ}(-3.5^{\circ})&\nonumber
 \eea
The results show that the positive (negative) $\delta_s$ leads to a positive (negative) shift of the mixing angle for $\mixing \lsim 45^{\circ}$, which is consistent to what we can observe in Fig.~\ref{fig:1}. And for this lower range  of $\mixing$, the uncertainties from the unknown $SU(3)$ effect could exceed the statistical error.
For 60$^{\circ}$ and 75$^{\circ}$, the trend of the sign of the shift is not seen and this is probably due to the large statistical error which causes a fluctuation, that is, the statistical error dominates over the $SU(3)$ breaking effect in this range of  $\mixing$. 
The bottom line is,  even after taking into account the systematic error coming from the $SU(3)$ breaking effect, on top of the statistical error, this measurement can still eliminate one of the two solutions, $\mixing\sim 30^{\circ}$ and $\mixing\sim 60^{\circ}$, obtained by the other experiments.

\section{Conclusions}
In this article, we proposed to measure the \fivebodydecay decay to determine the axial vector mixing angle $\mixing$. 
We first derived the \fivebodydecay differential decay rate formula in order to understand the $\mixing$ dependence of the $K\pi\pi\pi$ spectrum. {The theoretical formula for this} five body differential decay rate is obtained for the first time {in this article}. 
Using the obtained result, we performed a sensitivity study \comm{for determining  the $\mixing$ angle} by assuming the Belle (II) experiment environment. The $K\pi\pi\pi$ spectrum contains two $K_1$ resonances, $\kL$ and $\kH$. The contribution from $\kH$ diminishes as the $\mixing$ value increases. As a result, for a larger values of $\mixing$, let's say above $\sim 45^{\circ}$, the $\kH$ resonance becomes nearly invisible, which makes it difficult to distinguish the spectrums with different values of $\mixing$ in this range. More quantitatively, the expected statistical errors for 15k event are 
$\sigma_{\mixing}= \{\pm 1.3^{\circ}, \pm1.4^{\circ}, \pm1.3^{\circ},\pm 2.6^{\circ}, \pm 8.2^{\circ}\}$ for $\mixing=\{15^{\circ},30^{\circ},45^{\circ},60^{\circ},75^{\circ}\}$. 
This amount of data will be very soon available at the Belle (II) experiment. 

We also discussed a possible correction to this result due to the $SU(3)$ breaking effect, which is related to the production of $K_{1b}$ ($^1P_1$) state from the axial vector current. The existence of this contribution is not confirmed and we urge a theoretical progress on this matter. In order to evaluate its possible impact, we included the maximum of  $\pm$10\% $SU(3)$ breaking effect.  The result shows that it can shift the measurement of $\mixing$ by $\Delta\mixing= \{\pm 3.8^{\circ}, \pm 2.8^{\circ}, \pm1.4^{\circ}\}$ for $\mixing=\{15^{\circ},30^{\circ},45^{\circ}\}$. We find that the statistical error dominates in the case of the higher values of $\mixing$, i.e. $\mixing=60^{\circ} {\rm \ and\ } 75^{\circ}$. 

The other experiments found the $\mixing$ angle to be $\sim 30^{\circ}$ or  $\sim 60^{\circ}$ and to eliminate one of the solutions is a very important matter. For these values of $\mixing$, the \fivebodydecay can determine  $\mixing$ at the precision of  
\[\delta{\mixing}=\pm 3.1^{\circ} \quad {\rm (for\ \mixing=30^{\circ})},\quad  \delta{\mixing}=\pm 5.4^{\circ}\quad {\rm (for\ \mixing=60^{\circ})}\]
where the statistical uncertainty with 15k event at Belle (II) and the systematic uncertainty from 10 \% $SU(3)$ breaking effect are added by quadrature. Therefore, we conclude that the \fivebodydecay measurement can discriminate the two solutions for the $\mixing$ angle  obtained by the other experiments and determine it at this level of precision. 

\section*{Acknowledgement}
This work was in part supported by the TYL-FJPPL (France-Japan Particle Physics Laboratory). We acknowledge the Belle collaboration for letting us to use the simulation software. E.K. would like to thank F.~Le~Diberder and B.~Moussallam for valuable discussions.

\bibliography{arXiv_v2}{}
\bibliographystyle{unsrt}

\end{document}